\documentclass{iopjournal} 
\usepackage{amsmath} 
\usepackage{mathtools}
\usepackage{bbold}  
\usepackage[yyyymmdd]{datetime} 
\usepackage{tabularx} 
    \newcolumntype{L}{>{\raggedright\arraybackslash}X}
\usepackage{natbib}		
\setcitestyle{authoryear}	
\definecolor{lgreen}{rgb}{0.0,0.314,0.184}
\definecolor{blueviolet}{rgb}{0.2,0.05,0.90}
\hypersetup{allcolors=blueviolet}
\usepackage{braket} 
\renewcommand{\articletype}[1]{{\vspace*{-8mm}\noindent \large \sf arXiv version}\\[2ex]}

\begin{document}

\articletype{Topical Review} 

\title{Quantum annealing and condensed matter physics}

\author{V. Kendon$^1,*$\orcid{0000-0002-6551-3056} and N. Chancellor$^2$\orcid{0000-0002-1293-0761}} 

\affil{$^1$ Department of Physics, University of Strathclyde, Glasgow G4 0NG, UK}

\affil{$^2$ School of Computing, Newcastle University, Newcastle upon Tyne NE4 5TG, UK}

\affil{$^*$Author to whom any correspondence should be addressed.}

\email{viv.kendon@strath.ac.uk, nick.chancellor@newcastle.ac.uk}


\begin{abstract}
Quantum annealing leverages the properties of interacting quantum spin systems to solve computational problems, typically optimisation problems.  Current hardware now has capabilities that can be used to solve condensed matter physics problems, too.  In this \emph{topical review}, we provide an overview of quantum annealing aimed at condensed matter physicists, to show the mutual benefits of working together to understand and improve how quantum annealers work, and to use them to advance condensed matter physics.
\end{abstract}

\tableofcontents	
\section{Introduction}\label{sec:intro}

A quantum annealer is a device which computes by instantiating an Ising Hamiltonian on a set of quantum spins, with dynamics driven by a transverse field.  
A transverse Ising system is often the first example in textbooks on phase transitions in condensed matter systems \citep[e.g.,][]{Sachdev}.
There should, therefore, be much knowledge from condensed matter physics that can inform the development of quantum annealing, and in return, quantum annealers might be well-suited for simulating condensed matter physics problems.
This topical review aims to give condensed matter physicists insight into how the quantum annealing community views and exploits the physics of spin systems to solve a diverse range of problems.  It is not a comprehensive review of condensed matter applications of quantum annealing -- which would be far longer than space allows, given the number of different tests and experiments reported in the literature.  However, few of these claim an advantage over classical simulations, or produce new physics results.  Rather, they should be viewed as testing and developing suitable algorithms in readiness for more capable hardware, and in many cases, the algorithm itself is straightforward.  It is thus more relevant to focus on the current state of development of quantum annealers, both theory and experiments, and what might need to be improved to reach a realm where quantum annealing systems can be really useful for doing science.

We give a short overview of quantum computing in section \ref{sec:QCoverview}, to set quantum annealing in context.  Then, in section \ref{sec:QA} we provide definitions and more detailed descriptions of adiabatic quantum computing, computation by quantum walk, and quantum annealing, three closely related models of quantum computing.  In section \ref{sec:phys} we discuss how condensed matter physics informs how quantum annealing works, using two important examples.
In section \ref{sec:encode}, we discuss in more detail how to encode computational problems into quantum Hamiltonians, outlining the issues that arise from balancing the physical constraints with minimising the required resources.
To show how quantum annealers can be used for condensed matter physics simulations, in section \ref{sec:apps} we outline a few examples of how they have already been used, and 
we conclude in section \ref{sec:outlook} with some directions for future research, and thoughts on how quantum annealing is influencing the wider computing landscape.

\section{Quantum computing overview}\label{sec:QCoverview}

Quantum computing at its most general is a process of encoding a problem into a quantum state and applying a unitary transformation to this state, such that the transformed quantum state encodes the desired solution.  
The realisation that this can, in principle, solve certain problems more efficiently than classical computers \citep{Feyn1982,Deut1985,Feyn1986,Deut1992,Simo1994,Shor1994,Shor1995,Grov1996,Grov1997} founded the field of quantum computing.  
Forty years on, the practical challenges of turning this potential into reality are the focus of a vibrant sector of quantum technology development.

The most common model for quantum computing -- the quantum circuit model -- is a direct analogy with the classical circuit model underlying today's ubiquitous digital computers. 
The bits (binary digits) in classical computing are replaced by qubits (two state quantum systems) to which a unitary operation is applied, broken down into a sequence of discrete, standard one and two qubit unitary quantum gates.   
Similar to classical computing, there are quantum universal gate sets \citep{Bare1995} containing only a small number of elementary quantum operations, from which it is possible to construct any unitary transformation.  
The output of the computation is obtained by measuring some or all of the qubits, yielding a binary string, from which the solution is decoded.  
In some applications, the desired output is the quantum state of the register at the end of the process, rather than the classical bits obtained by measuring it.  
Strictly speaking, quantum state preparation is engineering rather than computation, but the two processes are closely related \citep{Hors2013}.

Qubits can be instantiated in any two-state quantum system that can be localised in space or time (photons) so that there are no identical particle statistical effects, and the individual qubits can be controlled and measured.
A diverse range of quantum computing hardware is under development, including superconducting (flux and current), silicon (electron spins), trapped ions, neutral atoms, photons, quantum dots, diamond vacancy centres.  
There is no clear leading design at this stage of development, for a recent overview, see \cite{QZeitgest2024hardware}.

A key ingredient in the progress towards practical quantum computers is \emph{error correction} \citep{Shor1995b,Knil1996,Ahar1996,Stea1996,Lafl1996} to suppress the effects of unwanted environmental interactions with quantum systems.  
In essense, this involves using multiple physical qubits to represent a single logical qubit, and using the extra degrees of freedom to detect and correct errors before they corrupt the computation.  A pedagogical review can be found in \cite{Roffe2019review}.

Roadmaps for quantum computing \citep[e.g.,][]{Acin2018roadmap} usually plot a path from current state-of-the-art noisy quantum processors to large-scale fully fault-tolerant, error-corrected digital quantum computers, the quantum equivalent of our digital silicon classical computer hardware. 
This is a logical goal, given the success and ubiquity of today's silicon CMOS technology.
However, multiple alternative models of quantum computing have been proposed and developed alongside digital quantum computing.  
Historically, this includes the original idea for quantum computing itself: simulating quantum systems \citep{Feyn1982}.  Using one quantum system to simulate the Hamiltonian dynamics of another quantum system was not given a digital setting until \cite{Lloy1996}. 
The more direct way to do quantum simulation is by engineering the Hamiltonian of another, easier to control, quantum system to match the system of interest.  The simulation evolves under the engineered Hamiltonian, without needing discretisation.  Properties of interest are then directly measured and translated back to the system of interest.

For other problems besides quantum simulation, the process of encoding into computational hardware requires a little more work to generate a faithful numerical representation of the problem.  Digital classical computers mainly use binary representations of integers and from these construct floating point numbers, the standard variable type for most computationally intensive problems.  Quantum computers currently only use binary integers, having not yet reached a size where floating point representations are practical.  There are a few other ways to encode data into quantum computers that are useful in specific settings, we discuss this in more detail in section \ref{sec:encode}.

Several models of quantum computing using qubits to represent data, but with a continuous time evolution under a Hamiltonian, were proposed relatively early in the development of quantum computing. 
Continuous-time quantum walk computation \citep{Farh1998}, adiabatic quantum computing \citep{Farh2000}, and quantum annealing \citep{Fini1994,Kado1998}, have all been developed theoretically and experimentally over the past decades, and are currently very active areas of research and implementation.   
Quantum annealing and special purpose quantum simulators, especially, are candidates for early quantum advantage, before the exacting demands of error-corrected digital quantum computing can be achieved.  In the next section, we describe quantum annealing and related models of quantum computing in more detail.  Other significant models of quantum computing, that are outside the scope of this review, include measurement-based quantum computing \citep{raussendorf01a}, and topological quantum computing \citep{PachosBook}.
Photonic models for quantum computing based on boson sampling \citep{Aaronson2011} are intermediate between classical and quantum in capabilities, leveraging the identical particle statistics of bosons.
Boson sampling is finding applications in quantum machine learning and neuromorphic models of computing, another very active area of research that is outside the scope of this review.
The same photonic hardware can also be used to solve optimisation problems 
\citep{bradler2021ORCA,goldsmith2024ORCA+TSP}, leveraging quantum optics rather than condensed matter physics.

\section{Quantum annealing and related models}\label{sec:QA}

Adiabatic quantum computing and computation by continuous-time quantum walk are variants of the same computational model using quantum evolution under the Hamiltonian 
\begin{equation}\label{eq:HABgen}
\hat{H}(t) = A(t)\hat{H}_0 + B(t)\hat{H}_p,
\end{equation}
applied to a set of qubits.
Here, $\hat{H}_p$ is a Hamiltonian encoding the problem, while $\hat{H}_0$ is a Hamiltonian that provides dynamics.  
The most common choice is the transverse Ising Hamiltonian, where $\hat{H}_p$ is the Ising Hamiltonian and $\hat{H}_0$ is the transverse field.  The qubits are instantiated as the spin-\textonehalf~particles to which the transverse Ising Hamiltonian is applied. Introducing some of our notation, it can be written
\begin{equation}\label{eq:HTI}
\hat{H}_{TI} = A(t)\sum_j \hat{X}_j 
             + B(t)\left\{\sum_j h_j\hat{Z}_j 
               + \sum_{j,k} J_{jk} \hat{Z}_j\hat{Z}_k\right\} 
\end{equation}
where the Pauli x and z operators applied to spin $j$ are denoted by $\hat{X}_j$ and $\hat{Z}_j$ respectively.  The field on the $j$th spin is $h_j$ and the coupling between spins $j$ and $k$ is $J_{jk}$.  The transverse field is usually applied globally to all the spins, so only controlled by $A(t)$.
The time-dependent real functions $A(t)$ and $B(t)$ determine how the time evolution proceeds.
The difference between quantum walks and adiabatic quantum computing is in how $A(t)$ and $B(t)$ are specified.  For quantum walks, they are constant throughout the time evolution, while for adiabatic quantum computing, there is a smooth transition of $A(0)=1$ to $A(t_f)=0$, while $B(0)=0$ to $B(t_f)=1$, for the duration of the computation $t_f$.  
The time dependence of $A(t)$ and $B(t)$ is chosen to keep the system close to the ground state.  
Quantum annealing is the same model but with open system dynamics playing a significant role.  The dynamics controlled by $A(t)$ and $B(t)$ is usually also faster than adiabatic, adding diabatic effects to the computational mechanisms, see  \cite{Crosson2020} for a more detailed recent review.

The most common way to encode problems into quantum Hamiltonians is to construct the Hamiltonian such that the solution to the problem is the ground state assignment of the spin variables.  
The unordered search problem provides the simplest example of how this works.
This is the problem of finding one item out of a set, where the only thing you know is that you will recognise the answer when you find it.  The items are labeled by binary integers, and the marked item has label $m$.  The Hamiltonian
\[
\hat{H}_s = \mathbb{1} - \ket{m}\!\bra{m}
\]
makes the marked state $\ket{m}$ one unit of energy lower than all the other states.  This is fine as a test problem, and has been widely studied
\citep[e.g.,][]{Shen2002,childs2003,Morley19interp,Berwald2024Zeno}.
However, constructing the Hamiltonian requires knowing the answer, so it is only useful as a test problem \citep{Dodds2019}, or as a subroutine in a larger computation, where the ``answer'' is identified as a condition from other parts of the calculation.
The Hamiltonian also is not Ising, the direct implementation requires higher order terms, although it can be implemented perturbatively in an Ising setting using gadgets and twice the number of qubits \citep{chancellor16a,chancellor17a}. 

Although it is, in principle, possible to encode any classical problem into an Ising Hamiltonian, a natural class of problems for Ising Hamiltonians is optimisation problems.  Many real world problems of significant commercial value are optimisation problems, so efficient methods of solution are useful and highly sought after.  It is possible to encode optimisation problems efficiently: \cite{Lucas2014} and \cite{Choi2011minor} provide encodings for many NP-hard optimisation problems.  We provide a more detailed discussion of encoding in section \ref{sec:encode}, after introducing the different methods for solving the problem.  For more background on quantum methods to solve optimisation problems, and the prospects for achieving useful applications, see the recent reviews \cite{Abbas2024,koch2025quantumoptimizationbenchmarkinglibrary}.

\subsection{Adiabatic quantum computing}\label{ssec:aqc}

Adiabatic quantum computing was first introduced by \cite{Farh2000}.  The setting is more general than a transverse Ising Hamiltonian, requiring only that the problem is encoded into an $n$-qubit Hamiltonian $\hat{H}_p$ such that the solution corresponds to the ground state of $\hat{H}_p$.  The qubits are initialised in the ground state of a simpler Hamiltonian $\hat{H}_0$ that is easy to prepare.  The full Hamiltonian, equation \eqref{eq:HABgen}, is then transformed slowly from $\hat{H}_0$ to $\hat{H}_p$,
\begin{equation}
    \hat{H}(t) = (1-s(t))\hat{H}_0 + s(t)\hat{H}_p. \label{eq:Haqc}
\end{equation}
The \emph{annealing parameter} $s(t)$
is monotonically increasing, and a function of the size of the problem space $N = 2^n$ and the \emph{accuracy parameter} $\epsilon$ determined by the adiabatic condition, 
\begin{equation}
    \frac{\Big| \Big\langle \frac{d\hat{H}}{dt} \Big\rangle_{1,0} \Big|}{(E_1 - E_0)^2} \equiv \epsilon \ll 1. \label{eq:adiabaticCond}
\end{equation}
Here $0$ and $1$ refer to the ground and excited states respectively, and
\begin{equation}
    \Big| \Big\langle \frac{d\hat{H}}{dt} \Big\rangle_{1,0} \Big| \equiv \bra{E_1}\frac{d\hat{H}}{dt}\ket{E_0}.
\end{equation}
By the quantum adiabatic theorem, for $\epsilon\to 0$ the qubits remain in the ground state throughout the transformation, hence the solution is represented by their final state, the ground state of $\hat{H}_p$.  
Note that we can make the Hamiltonian well-behaved by design, so it will remain gapped throughout the evolution, and the subtler aspects of adiabatic evolution in less well-behaved systems are not a concern.

The closer $\epsilon$ is to zero, the more completely the system will stay in the ground state and the longer the computation will take.  
The crucial part of the algorithm is thus how, and how slowly, the Hamiltonian must be evolved to ensure a high probability of remaining in the ground state. 
By theoretical-computer-science arguments, computationally hard problems should produce exponentially small gaps between the ground state and excited states at some point during the transformation, and thus the Hamiltonian must be evolved exponentially slowly to remain in the ground state. 
Were this not the case, an annealer (or a universal quantum computer simulating an annealer) could solve such problems in a time which is polynomial in the size of the problem. This would show the equivalence of P=NP when quantum methods are included, which is believed to be highly unlikely by computer scientists.  
Even for problems that are not exponentially hard, such as unordered search, an efficient adiabatic quantum algorithm needs a nonlinear function for $s(t)$ to provide a speed up over classical searching \citep{Rola2002}.  
This highlights the importance of careful control of the time-evolution for efficient computation.  
It also alerts us to the possibility that calculating the optimal control might be equivalent to, or harder than, solving the original problem.  
However, the adiabatic regime does allow a formal proof of (theoretical) quantum advantage in the unstructured search problem \citep{Rola2002}.  It can also be viewed as a specific implementation of a more general class of computations which make use of Zeno effects and were shown by \cite{Berwald2024Zeno} to support a similar formal advantage. Interestingly, \cite{Berwald2025Zeno} found that among Zeno mechanisms, the adiabatic setting appears to be the only one which allows frustration, and this seems to be an important consideration for practical computation.

While adiabatic quantum computing is very useful as a theoretical model, practical application in real hardware is limited by coherence times and noisy environments, and there is no practical route for scaling to useful sizes\footnote{Adiabatic state transfer is very useful in other settings, such as implementing quantum gates in the circuit model, \citep[e.g.,][]{Chancellor2012AdiabaticBus,Chancellor2013Holonomic}.}.

\subsection{Computation by quantum walk}\label{ssec:qw}

Computation by quantum walk was first proposed as a tool for quantum computation by \cite{Farh1998}. 
A continuous-time quantum walk is defined on an undirected graph $G$ with adjacency matrix $A$.  
The adjacency matrix has entries $A_{jk} = 1$ if (and only if) vertices $j$ and $k$ in $G$ are connected by an edge, and $A_{jk} = 0$ otherwise.  Since $G$ is undirected, $A$ is symmetric, so we can define a Hamiltonian based on $A$.  
The non-zero entries in $A$ specify the allowed transitions on the graph (edges), and the possible states of the system are the vertices of the graph.
It turns out to be most convenient to work with the Laplacian $\hat{\mathcal{L}}$ of the graph, defined as
\begin{equation}
    \hat{\mathcal{L}} \equiv A - D
\end{equation}
where $D$ is the diagonal matrix with entries $D_{jj} = \deg(j)$, the degree of vertex $j$ in $G$.  The Hamiltonian for the continuous time quantum walk on graph $G$ is then
\begin{equation}
    \hat{H}_w = -\gamma \hat{\mathcal{L}}
\end{equation}
where $\gamma$ is the hopping rate, and the quantum walk evolves according to the solution of the Schr\"{o}dinger equation, giving
\begin{equation}
    \ket{\psi(t)} = \exp(-i\hat{H}_w t)\ket{\psi_0}
\end{equation}
after a time $t$, starting in state $\ket{\psi_0}$, and where we have used units in which $\hbar=1$.

There followed several quantum walk algorithms exhibiting a quantum speed up, notably quantum searching \citep{Shen2002} and ``glued trees'' \citep{Chil2002}.  
Continuous-time quantum walks were proved to be universal for quantum computing by \cite{Chil2009}.  
The proof involves encoding the continuous-time quantum walk into a digital quantum computer, and showing that the Hamiltonian evolution can be efficiently approximated by a universal set of discrete unitary gates.  
While discretisation is rather unnatural for the Hamiltonian evolution, it is important to note that encoding the labels of the graph into qubits is critical for the proof of efficiency.  
An alternative, an \emph{architecture} for universal quantum computing based on multiple interacting quantum walkers has also been proposed by \cite{Chil2013}.  
As presented, both models of quantum walks are closely related to the gate model, with the graph structure resembling a circuit diagram.  They are thus less useful for developing continuous-time quantum computing using natural Hamiltonian interactions. 

\cite{Morley19interp} pointed out that continuous-time quantum walks can be framed as quantum annealing with a discontinuous schedule, with $A(t)$, $B(t)$ in equation \eqref{eq:HABgen} jumping from zero or one to a constant value at the start, and then to one or zero at the end.
Furthermore, \cite{Morley19interp} demonstrated that interpolations between continuous walks and adiabatic algorithms are possible and retain a quantum advantage, with intermediate protocols comprising a form of quantum annealing which does not fall into either category, the intermediate regime between diabatic and adiabatic in table \ref{tab:regimes}.
Mapping between quantum walks and adiabatic quantum computing allows results from one to be applied in the other setting, too.  For example, \cite{Omar2010random} determine the connectivity below which quantum walk search is no longer efficient in random graphs, and the same connectivity constraints can be expected to apply in adiabatic quantum search. 

There were few other applications of quantum walks to solve specific problems until \cite{Callison2019} showed that they can be used to find spin glass ground states.  Although the efficiency of the algorithm is better than a quantum search, it does not provide a speed up over classical branch and bound algorithms already known \citep{hartwig1984}, let alone the quadratically faster quantum enhanced version from \cite{Montanaro2015}.  Nonetheless, it does show that quantum walks can solve optimisation problems, and \cite{Mirkarimi2023} showed how they can potentially enhance portfolio solvers for SAT problems.  
Many important real world optimisation problems are NP-hard, and the best classical algorithms are at best a polynomial speed up over random guessing.  Quantum walks and quantum annealing often provide a quadratic speed up over classical, which is a very valuable improvement in a setting where exponential advantage is expected to be unobtainable by any method.

Most experimental implementations of quantum walks are physical walks, not encoded, and are therefore not directly useful for efficient quantum computing.  The quantum walk in the transverse Ising setting is taking place in Hilbert space, not real space, and can thus in principle be realised by the same hardware as quantum annealers.  In practice, the instantaneous ramps of the controls are not achievable, but the \emph{reverse annealing} capability in the D-Wave quantum system does effectively implement a quantum walk in the middle section where the controls are held constant \citep{Chan2016}.

\subsection{Quantum annealing}\label{ssec:qanneal}

In quantum annealing as first introduced by \cite{Fini1994} and \cite{Kado1998}, the initial state (which need not be the ground state of any Hamiltonian) is evolved under the problem Hamiltonian, relying on quantum tunnelling out of local minima and energy exchange with a low temperature environment to reach the ground state solution of the problem.  In other contexts, this setting is described as a \emph{driven dissipative system}.
Today, terminology is used more generically, and quantum annealing refers to quantum computational processes in which the Hamiltonian is evolved over time (annealing schedule, like $A(t)$ and $B(t)$ in equation \eqref{eq:HABgen}) as well as being subject to environmental coupling.  
The combination allows for a faster Hamiltonian evolution than a pure adiabatic strategy, with unwanted excited states able to lose energy to the environment, reducing their amplitude.  
Experimental evidence suggests that, until the recent development of fast annealing \citep{King2022Coherent}, this energy loss mechanism played a dominant role in the D-Wave computational devices \citep{Dick2013}. 
Quantum annealing works well when the energy landscape has many closely-spaced false minima, through which quantum tunneling is efficient.  Wide energy barriers slow down the tunneling. 
However, hybrid strategies can be devised to evade wide minima in cases where these are significant \citep{Chan2016,Chan2017}.
Evidence for a scaling advantage for an approximate form of quantum optimisation has been shown, but only on problems which match the native hardware graph, or are derived from the native graph through applying error correction \citep{Munoz-Bauza2025QAadvantage}.  Large enough problems to attain an actual advantage over classical in this setting have not yet been demonstrated.

Quantum annealing running on real hardware manifests several distinct regimes, each with their own behaviour that can be controlled through the annealing schedule equation \eqref{eq:HABgen}, specified by $A(t)$, $B(t)$, and the total anneal time $t_f$:

\subsubsection*{Adiabatic regime:} 

In this regime, open system effects can be neglected, and time evolution is slow enough for the system to remain in the ground state with very nearly unit probability for the entire evolution. Adiabatic quantum computing is quantum annealing performed within the adiabatic regime. While conceptually easy to understand, because it can be reduced to a two-state subspace of ground and excited states, this regime is unlikely to be relevant for real problems, especially for NP-hard optimisation problems where the required runtime must be exponentially long (see section \ref{ssec:aqc}).

\subsubsection*{Quasistatic regime:}

In this regime, the thermal dissipation timescale is short enough relative to other timescales for the system to be treated as being in thermal equilibrium.  Quasistatic dynamics is well described by an approximate Kibble-Zurek mechanism where the dynamics stops suddenly as the equilibration timescale becomes longer than the timescale of the anneal \citep{Kibble1976KZ,Zurek1985KZ,chancellor2016DWaveKZ,Gardas2018DWaveKZ,Weinberg2020DWaveKZ,Bando2020DWaveKZ}. To a good approximation in many cases, evolution in the quasistatic regime allows for thermal sampling \citep{Benedetti2016Temperature}, which is a valuable tool in statistical inference \citep{Chancellor2016MaxEnt,Benedetti2017PGM} and machine learning \citep{Adachi2015Boltzmann,Amin2018Boltzmann,Benedetti2018Helmholtz,Caldeira2019Botlzmann,Sinno2025Boltzmann}. The effective sampling temperature is considered an important metric for quantum annealers \citep{Lall2025Metrics}. While this sampling isn't perfect, it is still potentially useful, and produces intriguing phenomena, for example the ``global warming'' effect reported by \cite{Raymond2016warming}, where larger scale measures of temperature yield higher values. Interesting statistical mechanical effects can also be observed, for example, the counter-intuitive noise effects detailed by \cite{Chancellor2022Error}, which can be used to measure control errors in the underlying device. An additional question related to this kind of sampling is whether or not annealers can fairly sample ground-state manifolds.  The ``fair sampling'' problem has been studied extensively, for example in \citep{Konz2019FairSample,Kumar2020FairSample}. While unfair sampling is potentially problematic in some use cases, it has been suggested by \cite{Zhang2017UnfairAdvantage} that it could be used to complement other methods, and if done carefully, the unfair nature could be useful.  Furthermore, \citep{Chancellor2020FluctGuided} suggested that this unfair sampling will tend to favour solutions to optimisation problems with useful properties. 

It is an open question whether the type of thermal sampling provided by quantum annealers -- or, indeed, any type of thermal sampling -- can provide a computational advantage.  Recent work has found that quantum annealing has a scaling advantage over the classical state-of-the art for approximate optimisation \citep{Munoz-Bauza2025ApproxAdvantage}, and this suggests that the related problem of thermal sampling is a good general direction to seek out an advantage. It is important to note that an advantage has not yet been found for exact optimisation, suggesting that thermal sampling of low lying states, rather than ground-state sampling, may be the best place to look.

\begin{figure}[!hbt]
    \begin{center}
	    \includegraphics[width=0.7\columnwidth]{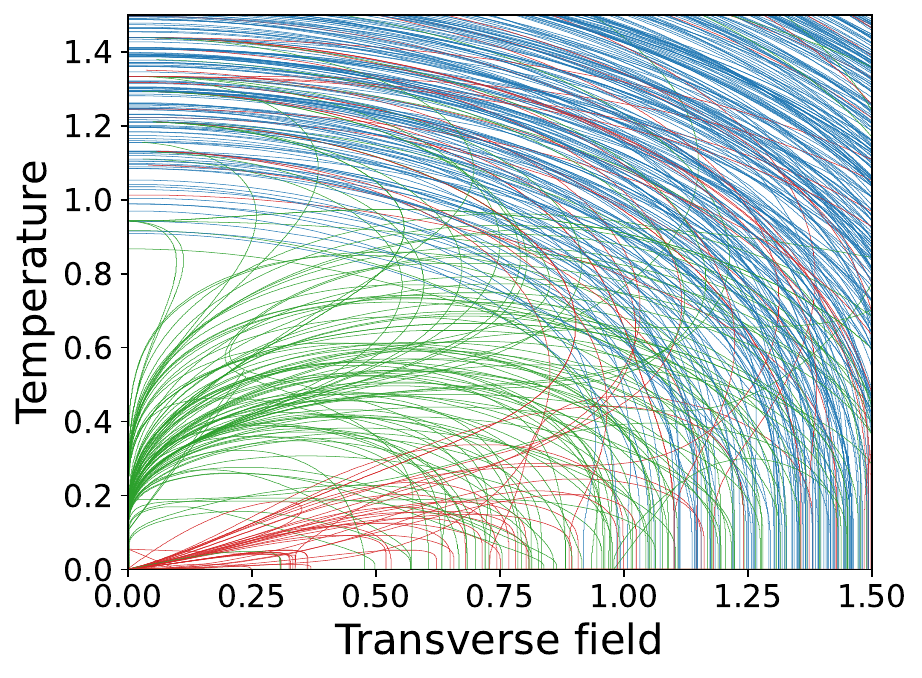}
	\end{center}
	\caption{Illustration of changes in the average sign of spin values versus temperature and transverse field for 3x3 square spin glasses. This plot represents the places where spins change sign in all possible combinations of ferromagnetic and anti-ferromagnetic edges on this lattices. In some cases, spins react the same way to temperature and transverse field (blue), while in others (green and red) they illustrate complex behaviour. These categories can further be determined based on behaviour at low temperature and transverse field, as explained in \cite{chancellor2016DWaveKZ}.  Figure adapted from figure 5 of \cite{chancellor2016DWaveKZ}, which also uses this system to analyse frozen-in fluctuations on a superconducting flux-qubit quantum annealer operating in the quasistatic regime. \label{fig:QA_spin_flips}}
\end{figure}

The ability to sample thermally also provides an opportunity to study equilibrium condensed matter systems. Flux qubit devices with long anneal times naturally operate in this regime, but it may be difficult to access with atom or ion based systems as they are not in contact with a finite-temperature thermal bath, they are cooled by driving specific  transitions with lasers \citep{Wineland1978}. It has also been found that, in some cases, the effect of a frozen-in transverse field can mimic thermal fluctuations, while this picture fails in other cases \citep{Otsubo2012Thermal,Otsubo2014thermal,chancellor2016DWaveKZ}, so it is unclear to what extent a regime dominated by quantum rather than thermal fluctuations can be useful for approximate thermal sampling. Note that while coupling to the thermal bath dominates the dynamics, this regime is still highly quantum, and numerous papers have found evidence of quantum effects within this regime \citep{Johnson2011Manufactured,Lanting2014Entanglement}. These frozen-in effects were studied in \cite{chancellor2016DWaveKZ}, for example, and even at equilibrium there is a complex interplay between transverse field and temperature, as illustrated in figure \ref{fig:QA_spin_flips}.

\subsubsection*{Diabatic regime}

In this regime, the annealing timescales are much too short for the evolution to be adiabatic, and also much shorter than the timescales relevant to thermal coupling. Mathematically, the evolution within this regime is well-described by unitary time evolution, but direct, full-state-space simulations are not feasible for the sizes of current hardware. In fact, recent experiments on flux qubit annealers could realise the model better than it could be simulated even by the most advanced tensor network simulations \citep{King2025beyondClassical}. This regime can be accessed by fast annealing on a flux qubit device and is also likely to be relevant to atom and ion systems. Understanding the mechanisms at play in this regime is an area of active research with a number of recent papers proposing ways in which it can be understood. One way is by conceptualising the dynamics as a series of very short quantum walks, where the hopping is reduced at each stage \citep{Callison2021,Banks2024continuoustime,Schulz2024,Gerblich2024}. Since quantum walks conserve energy and it has been shown \citep{Callison2021} that the reduction of the hopping parameter can only take energy out of a system, this implies that diabatic annealing must act as an energy-removal mechanism. Related, but distinct, ways to understand these systems include through Planck's principle \citep{Banks2025Diabatic}, or by invoking ideas from optimal state transfer \citep{Banks2024rapidquantum}.

With short anneal times and high coherence, the diabatic regime is an obvious place to look for a possible
advantage in transverse Ising quantum annealing. Optimal control theory implies an optimal annealing schedule exists \citep{Brady2021,Gerblich2024}, while finding it is effectively solving the problem another way.  There is thus a delicate balance between approximating optimal annealing schedules well enough to be useful, and transferring the computational effort over to calculating the schedule.  A good example of how this balancing works is given by \cite{Schulz2024}.  

A further diabatic method involves neutralising the unwanted effects of the fast dynamics by adding extra terms to engineer counterdiabatic driving \citep[e.g.,][]{Cepati2023}.  Counterdiabatic terms are usually of a form that goes beyond a transverse Ising Hamiltonian in a way that creates a total Hamiltonian that has been shown to be equivalent to universal quantum computing \citep{Aharonov2004} when combined with adiabatic evolution.  This puts the question of a computational advantage into a different setting, since universal quantum computing has proven speed ups for specific problems.  The transverse Ising Hamiltonian by itself is not universal for quantum computing, the Ising Hamiltonian can only efficiently encode classical problems.  Counterdiabatic terms are hard to engineer in real hardware and are not currently available in superconducting quantum annealers. However, in neutral atom hardware, counterdiabatic terms should be possible in principle, given that much of the development of neutral atom hardware is focused on gate based universal quantum computing.

\subsubsection*{Intermediate regimes} 

For hardware with significant open system effects, there must also be an intermediate regime between the diabatic and quasistatic regime, where open system effects cannot be ignored, but also do not dominate. Such a regime will be even harder to understand conceptually, but, due to the limited time spent in it, it is unclear how important this regime is algorithmically. Similarly, for systems with longer coherence times such as neutral atoms, there will be a smooth transition between diabatic and adiabatic regimes, as the adiabatic approximation becomes more accurate. 

Not all hardware will be able to reach every regime. The ability to reach the adiabatic regime is going to be highly dependent on the Hamiltonian being simulated, as it depends crucially on the minimum gap. This means that for spin-glass like Hamiltonians, where the minimum gap scales exponentially, the adiabatic regime is likely to only be accessible for small problems, since the exponential scaling is likely to make the inverse gap scale rapidly beyond any practical timescale. For other Hamiltonians, the same arguments may not apply as the scaling could be much weaker. 

\subsection{Current hardware capabilities}

For the definition we are using, accessing the quasistatic regime is only possible in systems where the dominant error is from interaction with a finite-temperature thermal bath. Superconducting flux qubits do have this property: the presence of a thermal bath with quasistatic operation makes these devices useful for thermal sampling \citep{Chancellor2016MaxEnt,Benedetti2018Helmholtz,Nelson2022Sampling,Sandt2023Sampling,Sinno2025Boltzmann}, and the distribution temperature is a useful metric for performance \citep{Lall2025Metrics}. Current state-of-the-art flux-qubit devices have been demonstrated to be able to access both the diabatic and quasi-static regimes \citep{King2022Coherent}. Hypothetically, they should be able to operate adiabatically on small and simple Hamiltonians, but we are not aware of any such experiments\footnote{This is likely due to the fact that such experiments would contribute little to scientific knowledge, not because they could not be achieved for technical reasons.}. However, in neutral atom systems, while there are potentially some sources for thermalisation from the laser trapping potential, atom loss is currently the dominant error \citep{Ma2023Rydberg,Saffman2016Rydbergreview}. In such systems, atoms would be lost before they could thermalise, and therefore the quasistatic regime cannot be reached. This leaves only the adiabatic and diabatic regimes for neutral atom systems, with similar caveats for how useful the adiabatic regime is likely to be for larger problems. 

The diabatic regime should be reachable by any system with sufficiently low noise, and therefore could be argued to be the most important regime to study. It is also the least understood. In principle, as an annealer becomes large enough, it may act as its own thermal bath \citep{Linden2009}, and a quasistatic approximation may become accurate. Understanding if, how, and at what sizes this happens for different systems is a potentially interesting research question by itself, and one where some interesting experimental \citep{Ueda2020thermalisation}, and theoretical progress \citep{Gogolin2016Thermalisation,Deutsch2018EigenstateThermalisation} has been made. While we are not aware of such research being performed on annealing hardware, it certainly could be in the future. 

To formalise the different regimes, we can compare relevant rates to coupling constants. For this discussion we assume dimensionless units. Firstly, we define the annealing rate $\frac{\partial \Gamma}{\partial t}$, where $\Gamma(t)=\frac{A(t)}{B(t)}$ with $A(t)$ and $B(t)$ being the prefactors defined in equation \eqref{eq:HABgen}. Note that, because of the importance of non-linear schedules \citep{Rola2002} in obtaining a quantum advantage in adiabatic annealing, it is not appropriate to simply use the inverse of the overall annealing time here. Secondly, the relevant energy scale for adiabaticity is defined by the minimum value the gap $\Delta=\min(E_1-E_0)$ takes. Lastly, we define the coupling coefficient to an external bath $\gamma$. By comparing these values at the point where the gap is minimum, we can mathematically establish characteristics which define regimes. For example, if $\gamma\gg \frac{\partial \Gamma}{\partial t}$, then we would expect thermal equilibration to dominate and therefore to be in the quasistatic regime.  On the other hand, if $\gamma\ll \frac{\partial \Gamma}{\partial t}$ and $\Delta\gg \frac{\partial \Gamma}{\partial t}$, we would be in the adiabatic regime. A full summary of the relationships defining these regimes is given in table \ref{tab:regimes}.

\begin{table}[!hbt]
    \centering
    \begin{tabularx}{\linewidth}{|L||L|L|L|}
     \hline   
     Relationships & $\gamma\ll  \partial \Gamma/\partial t$   & $\gamma\approx \partial \Gamma/\partial t$ & $\gamma \gg \partial \Gamma/\partial t$  \\ \hline \hline
       $\Delta \gg \partial \Gamma/\partial t$  &  adiabatic & intermediate adiabatic/quasistatic & quasistatic \\ \hline
       $\Delta \approx \partial \Gamma/\partial t$ & intermediate diabatic/adiabatic & intermediate adiabatic/diabatic/\-quasistatic & quasistatic \\ \hline
       $\Delta \ll \partial \Gamma/\partial t$ & diabatic & intermediate diabatic/quasistatic & quasistatic \\ \hline
    \end{tabularx}
    \caption{Regimes of quantum annealing defined in terms of relationships between annealing rate and various parameters: $\partial \Gamma/\partial t$ is the annealing rate, $\gamma$ is the coupling to a thermal bath, and $\Delta$ is the minimum gap. Note that this table assumes dimensional scaling so that energy is equivalent to inverse time.}
    \label{tab:regimes}
\end{table}

The regimes summarised in table \ref{tab:regimes} provide useful insights into the different protocols.  In particular, because of the effectively instantaneous parameter changing at the start and end of the ``anneal'', quantum walks only ever operate in the diabatic regime, while adiabatic quantum computing is, by definition, in the adiabatic regime. It may also be useful to understand the three regimes which are intermediate between adiabatic and other regimes, for example \cite{Venuti2016AdiabaticOpen} examine how concepts of adiabaticity can be extended to open systems. We have listed the annealing methods along with their operating regimes and defining characteristics in table \ref{tab:algorithms}.

\begin{table}[!htb]
    \centering
    \begin{tabular}{|c||c|c|c|}
     \hline   Anneal Method & Operating Regime & Defining Characteristic  \\ \hline
    \hline   Adiabatic Quantum Computing & Adiabatic & Adiabatic Condition e.g.~Eq. \eqref{eq:adiabaticCond}   \\ \hline 
    Quantum Walk & Diabatic & $A(t)$ and $B(t)$ constant (except $t=0,t_f$) \\ \hline
    Multi-Stage Quantum Walk & Diabatic & $A(t)$ and $B(t)$ stepwise constant \\ \hline
    Quantum annealing & Any & Hamiltonian of form Eq.~\eqref{eq:HABgen} \\ \hline
    \end{tabular}
    \caption{Table categorising the quantum annealing methods we have discussed in terms of their operating regimes (see table \ref{tab:regimes}) and defining characteristics.}
    \label{tab:algorithms}
\end{table}

An interesting question is whether annealing like-protocols are optimal in a digitized setting or whether an entirely different approach can perform better.  Formally, it was shown from optimal control theory that bang-bang control sequences, where a system alternates between extreme parameter values, are optimal due to the Pontryagin's minimum principle \citep{Zhi-Cheng2017bang-bang}. However, such formal solutions include unphysical controls, where the values are changed an infinite number of times within a finite interval \citep{Borisov2000Fuller}. This left the question open as to whether annealing like protocols are optimal within the physically implementable ones. \cite{Brady2021} showed that indeed for physically realisable protocols the optimal limit of a discretised form of quantum annealing suitable for digital quantum computers is a smooth quantum annealing schedule \citep{diezvalle2025universalresourcesqaoaquantum}. Furthermore, \cite{Gerblich2024} showed that multi-stage quantum walks are a better approximation for quantum annealing than alternating applications of the driver and problem Hamiltonians, as first proposed by \cite{farhi2014QAOA}.
This tells us that the alternating driver and problem Hamiltonian approach on digital quantum hardware cannot improve on quantum annealing in an ideal setting, but leaves open the possibility that there is a different approach with better performance.

\section{Condensed matter physics of quantum annealing}\label{sec:phys}

An important reason for applying condensed matter physics theory to quantum annealing is that the kinds of questions it can answer tend to be complementary to those answered by traditional computer science techniques. Computational complexity theory tools are largely built on the concepts of reductions \citep{Karp1972Reduce}, in other words mapping of one problem to another. This means that computational hardness results, (e.g., NP-hardness), tend to be worst case results, with implications about what instances can exist, but saying nothing about how average case, i.e., typical instances, behave. This distinction is important: there are well known examples of NP-hard problems where typical instances are easy, for example, knapsack problems \citep{Beier2004KnapsackEasy} and some satisfiability problems \citep{krivelevich2006SatisfiabilityEasy}. In the case of knapsack problems, in which the problem is to exactly fill a container with smaller items from a given set, it is even challenging to construct sets of difficult instances \citep{Jooken2022HardKnapsack}. In fact, many classic video games are \emph{formally} NP-hard \citep{Aloupis2015Nintendo}, but \emph{typical} levels do not require players to have advanced optimisation algorithm solving skills.

In contrast, many tools of condensed matter physics are built around understanding thermal behaviour, and therefore aim to capture typical instances and behaviours. A concrete example is spin glass theory, which captures the typical behaviour of problems with inherently random structures. This perspective can be valuable even for benchmarking in a more traditional optimisation setting. This was highlighted by early benchmarks of quantum annealers which were based on random couplings using the hardware's native quasi-planar architecture \citep{Ronnow2014Detecting}. When viewed from a computational complexity theory perspective, this seemed like an effective way to benchmark. Reductions exist to show that Ising problems even on such a restricted graph are NP-hard \citep{Barahona1982IsingHard}. However, when viewed from a spin-glass perspective, an important issue became clear: these models only exhibited a spin-glass transition at zero-temperature, meaning that typical cases would be relatively easy for classical Monte-Carlo algorithms \citep{Katzgraber2014ChimeraBlind}. In this case, not only did spin glass theory prove to be a valuable tool for diagnosing an issue in a benchmarking proposal, it became a valuable tool in designing better benchmarks \citep{Katzgraber2015SpeedupSpinGlass}.

\subsection{Quantum Monte Carlo methods}

\begin{figure}
    \begin{center}
	    \includegraphics[width=0.6\columnwidth]{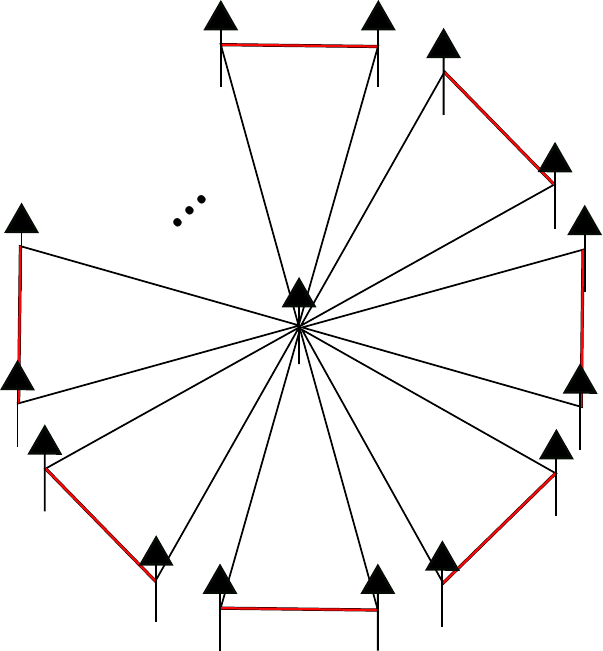}
	\end{center}
	\caption{Illustration of the ``shamrock'' Hamiltonian used by \cite{Andriyash2017QMC} to show an exponential separation between quantum Monte Carlo and physical, incoherent annealing. Black edges are ferromagnetic couplings of unit strength, while red are anti-ferromagnetic and of slightly weaker strength. The arrows show the configuration of the spins in one ground state. Ellipses (...) indicate there are many more ``leaves''.\label{fig:shamrock}}
\end{figure}

Other tools from condensed matter physics have also been fruitfully applied within a quantum annealing setting, increasing understanding within both settings. Quantum Monte Carlo\footnote{The form of quantum Monte Carlo we are discussing here is a completely classical algorithm, despite the potentially misleading name. Sometimes quantum Monte Carlo has been used to describe the use of quantum computers to aid in Monte Carlo calculations, but that is not the topic of discussion here.} (QMC), particularly path-integral-QMC techniques, have been used to emulate quantum annealing by slowly varying the transverse field parameter  \citep{Martonak2002PIQA}. Initially, such simulations were used to provide evidence of a potential advantage from quantum annealing \citep{Santoro2002QA}. While helping to motivate the field, this also raised an important question: \emph{does physical annealing of Ising models in the quasistatic regime present any advantage over QMC simulations?} 
Given that QMC only describes equilibrium distributions, it is fairly obvious that QMC cannot effectively describe diabatic annealing (see section \ref{ssec:qanneal} for regime descriptions).
It turns out that there is an advantage from physical annealing even outside of the diabatic regime, but this answer is nuanced and took many years to find. For simple tunneling, QMC simulations match what is expected from quantum theory \citep{Isakov2016QMCTunnel}, suggesting that if an advantage were present, it would have to come through a more complex mechanism. Conceptually, QMC can capture individual tunneling rates, but does not capture the effects of interference between different paths, a key effect in real quantum systems. Formally, this leads to what are known as topological obstructions -- features in the landscape which cause QMC to fail \citep{Hastings2013Obstructions}. As it turns out, such situations arise naturally in frustrated spin systems.  \cite{Andriyash2017QMC} demonstrated a simple ``shamrock'' geometry of spins, as shown in figure \ref{fig:shamrock}, for which physical annealing exhibits exponentially faster tunneling. This faster rate is due to the presence of exponentially many possible tunneling paths. Not only was this shown theoretically, it was later demonstrated experimentally \citep{King2021FrustQMC}. While QMC cannot fully emulate physical annealing, it is still possible that it could provide a performant quantum-inspired algorithm in suitable settings \citep{Lee2000QMCInspired,Zhang2021QMCInspired,Forno2025QMCInspired}.

\subsection{Physics of the diabatic regime}
A key challenge which is emerging in quantum annealing is understanding the diabatic regime. Fundamentally, the challenge is understanding far-from-equilibrium condensed matter physics. In simple systems, the quantum theory itself can provide a tool \citep{King2022Coherent}. However, in larger more complex systems, like those which encode real computational problems, classical simulations would be intractable, even with advanced methods such as tensor networks \citep{King2025beyondClassical}. All techniques which do not rely on directly solving the quantum dynamics, at least implicitly assume that the system is close to equilibrium: the Kibble-Zurek mechanism assumes sudden freezing of an equilibrium system; quantum Monte-Carlo techniques rely on thermal equilibration to do updates; and the adiabatic picture assumes a system is approximately in the ground state.  The latest results from D-Wave quantum annealers imply they are solving problems we cannot fully solve classically \citep{King2025beyondClassical}.  However, the problems are not yet useful apart from showing what the hardware is capable of: claims of quantum advantage or quantum utility would be premature.  In the meantime, this presents an exciting opportunity: the need for new theory to understand the next generation of quantum annealers, and the possibility of using current hardware to discover and test this theory. For the condensed matter physics community, such theory would be more widely applicable to problems of current interest, providing motivation to collaborate.

\section{Encoding problems into quantum Hamiltonians}\label{sec:encode}

The subtleties involved in how to encode optimisation problems into spin Hamiltonians are worth further discussion, since the configuration of the spins determines how they can be used.  
The fact that finding Ising ground states is NP-hard implies, through the Cook-Levin theorem, that at least in a very formal sense, any NP-hard problem can be reduced to it \citep{Cook1971NP,Karp1972Reduce}. 
However, to be useful, it is important to do any such reduction efficiently. 
In practice, this means using more direct mappings rather than the formal methods reductions used in proof.  \cite{Lucas2014} and \cite{Choi2011minor} provide efficient encodings for many NP-hard problems, like \emph{3SAT} and \emph{maximum-weight independent set}.
These are still not directly useful real world optimisation problems, but the methods can be straightforwardly adapted. 

\subsection{Hardware graph connectivity}
An important issue which arises in problem mapping is the hardware graph connectivity. For matter-based systems, such as superconducting qubits or neutral atoms, the hardware graph inherits the underlying two- or three-dimensional connectivity of the physical system, in the sense that the qubits (or qudits -- $d$-dimensional quantum systems) can only interact within a local neighbourhood in space%
\footnote{Neutral atom Rydberg systems do have long range interactions, but the scaling with distance is $1/r^6$, so rapidly becomes negligible, except for accounting for errors.}. 
This presents a problem in the traditional optimisation setting, where problems often have long range connectivity. 
This is fundamental: integer linear programming \citep{Rardin1998OpResearch}, which is often used in operational research, builds on equality constraints of the form 
\begin{equation}
    \mathbf{A}\vec{x}=\vec{b}
\end{equation}
where $\vec{x}$ is a vector, $\mathbf{A}$ is a constraint matrix and $\vec{b}$ is a vector of constraint values. When converted to a quadratic expression, this takes a form 
\begin{equation}
    \left(\mathbf{A}\vec{x}-\vec{b}\right)^T\left(\mathbf{A}\vec{x}-\vec{b}\right)=\vec{x}^T\mathbf{A}^T\mathbf{A}\vec{x}-2\vec{b}^T\mathbf{A}\vec{x}+\vec{b}^T\vec{b}
\end{equation}
which becomes fully connected if $\mathbf{A}^T\mathbf{A}$ is dense. 

While mappings are known that take fully connected problems to lower dimensional settings \citep{Choi2008minor,Choi2011minor,Lechner2015LHZ,Palacios2025Encode}, there are fundamental limitations on the number of physical units that need to be used to map a variable in the original problem. 
This can be understood rigorously using graph-theoretical tools such as treewidth \citep{Diestel2025Graph}. 
Conceptually, the graph needs to be ``wide'' enough to share information across all the variables (qubits). In physics terms, this implies that the surface area of one part of the graph should scale as the number of qubits in a fully connected graph.  This means that to map to a quasi-two-dimensional system, the number of computational units (spins) scales as the square of the number of variables in highly-connected problems. 
For a quasi-three-dimensional setting, the number will scale slightly better, as the power of $\frac{3}{2}$. 
This polynomial increase in the number of variables needed to map a problem to practical hardware is a barrier for scaling to useful problem sizes. 
To account for this, it has recently been suggested that, at least in the context of optimisation, a more appropriate metric to compare quantum annealers is the largest mappable fully-connected problem \citep{Lall2025Metrics}, rather than the raw number of qubits. 

A more serious issue in mapping a single variable to multiple spins is that constraints are needed to ensure the mapping is faithful. 
The simplest kind of constraints are known as \emph{minor-embedding} constraints \citep{Choi2011minor}, where multiple spins are strongly ferromagnetically coupled to represent a single logical variable. This can be seen by the frequency of chain break events when performing minor-embedded benchmarking \citep{Pelofske2025MinorEmbedBenchmark}.
If single spins are flipped directly, flipping a whole variable requires transitioning through high energy states, greatly slowing down the dynamics.  While there have been recent proposals to implement logical operations that avoid these issues \citep{Headley2025gadget}, they still represent a significant problem for scaling to useful sizes.

\subsection{Unary vs binary encoding}
The discretisation using bits or qubits to represent binary numbers is not the only way to encode data into physical devices. 
Like their classical counterparts, analog quantum computers use continuous variables to encode the problem \citep{Lloy1999,Brau2004}, with a different universal set of discrete, unitary transformations to evolve the quantum state.  
A physical quantity that can be varied continuously, such as length, position, or momentum, is used to approximate a real number, by making it proportional in size.  
However, for classical data, a unary encoding such as this is in general exponentially inefficient compared with a binary or digital encoding. 
Using $n$ bits we can represent $2^n$ different numbers, exponentially more than the number of bits.  Analog encoding is proportional to the size of the quantity.  A number twice as large doubles the size of the analog quantity representing it, whereas it requires only a single additional bit for binary encoding. 
Efficient classical computation therefore requires binary data encoding, and this is the main reason why digital quickly replaced analogue in the last century.  
The same efficiency arguments apply to encoding classical data into quantum computers, where the most common choice will be qubits, two-state systems.
It doesn't need to be binary, any reasonable choice of qudit of dimension $d$ will do \citep{Blum2002,Gree2004}.
There are implications for the complexity of the gate operations in digital quantum computing for $d>2$, but for problems where the variables are naturally trinary, for example, the extra complexity may still provide a net gain in simplicity.

For quantum annealing, however, there is an extra trade off to consider: the set of interactions which are allowed between binary-encoded variables are restricted to be quadratic.  In other words, for variables $V_1$ and $V_2$, the only interaction allowed is of the form $V_1\times V_2$.  Likewise, only single-body terms of the form $V_1$ and $V_1^2$ are possible. 
Such a quadratic structure is natural in some computational problems, such as factoring, but higher order terms are helpful in many practical optimisation settings.  
These quadratic restrictions don't apply to discrete unary encoded variables. In a unary encoding, interactions can select any of the possible values for each variable, allowing for arbitrary pairwise interactions to be represented as quadratic terms in the problem Hamiltonian. A simple degree-of-freedom-counting argument shows that such arbitrary interactions cannot be produced purely from quadratic and linear terms in denser encodings. In a binary encoding, the number of bits needed to represent an integer $N$ scales as $\log(N)$. The number of quadratic interaction terms between bits which are possible is therefore proportional to $\log(N)^2$.  However, representing arbitrary interactions (equivalent to arbitrarily high-order polynomials in the two variables), requires $N^2$ degrees of freedom.  In binary encoding, $N^2$ interactions therefore requires higher order terms, while $N^2$ is the scaling of quadratic degrees of freedom for a unary encoding. The exact nature of these tradeoffs has been explored extensively in \cite{chancellor2019,Chen2021DomainWall,Berwald2023Domain}. Moreover, when a variable is restricted to a finite range, unary encoding can be efficient, with only a constant encoding overhead offset by the simplicity of the interactions. Unary encoding strategies for individual variables are thus important in a quantum annealing setting where more general interactions are helpful.

\subsection{One-hot and domain-wall encodings}
An example of a unary encoding that works well in a spin system is a \emph{one-hot encoding}, which is commonly used in linear programming, predating quantum annealing \citep{Rardin1998OpResearch}. A one-hot encoding in spins is achieved by taking $n$ spins and enforcing quadratic constraints such that only one of them can take a $1$ value.   Early annealing examples include \cite{Venturelli2016JobShop} and \cite{Stollenwerk2019gate,Stollenwerk2020flight}.
An alternative technique, inspired by condensed matter physics, is a \emph{domain-wall encoding}, where each value is represented as the position of a domain-wall topological defect \citep{chancellor2019}. This representation requires one fewer binary variable per higher-than-binary variable than one-hot, and by general counting argument achieves the best theoretically possible density for arbitrary interactions \citep{Berwald2023Domain}.
Both representations allow for arbitrary interactions between variables, not just quadratic.
Experimentally, domain-wall encoding has been shown to be the superior method on D-Wave systems, leading to a later freeze time \citep{Berwald2023Domain}. Better experimental performance was also demonstrated on an older generation of flux-qubit annealers, where domain wall encoding could out-perform one-hot encodings of the same problem on more advanced hardware \citep{Chen2021DomainWall} .
Domain wall encoding has also been applied to simulating quantum field theory \citep{Abel2021QFT} on quantum annealers.

\section{Quantum annealing applications in condensed matter physics}\label{sec:apps}
%
Although the primary drive for developing quantum annealing hardware is to solve optimisation problems, this is not necessarily where the first benefits will be found.  Table \ref{tab:applications} summarises the current state-of-the-art and near term prospects for useful applications.  At current hardware sizes, useful physics research advances are only possible for niche problems well-matched to the hardware configurations.  There are many proof-of-concept examples in the literature that give a good indication of the potential as more advanced hardware becomes available.  We give only a few examples here, since our main focus is to make the concepts driving quantum annealing development more accessible to condensed matter physicists.  
Quantum utility is a moving target: classical algorithms are being improved as well, in part spurred by overambitious claims by quantum hardware developers.  Tensor network methods \citep{Berezutskii2025} are the most important for simulating quantum computing, and are finding multiple uses beyond their original application to condensed matter simulation.

\begin{table}[!htb]
\begin{tabularx}{\linewidth}{L|L|L}
\textbf{Application area} & \textbf{Current status} & \textbf{Near term prospects}\\
\hline
Optimisation problems & Many proof-of-concept demonstrations on real hardware.& Niche applications may be useful soon, but hardware graph connectivity is a major issue.\\
\hline
Other classical problems & Proof-of-concept demonstrations, e.g., quantum field theories, factoring. & Significantly larger, and in some cases more highly connected, hardware needed for advantage.\\
\hline
Equilibrium problems & Multiple demonstrations at beyond classical scale for specific well-chosen problems. & Potentially useful for interesting problems soon.\\
\hline
Out of equilibrium problems & Demonstrations, e.g., dynamics around phase transitions. & Potentially useful now for problems that map straightforwardly to available hardware graphs. \\
\hline
Hybrid quantum-classical algorithms (any area) & Proof-of-concept workflows demonstrated. & Likely useful as applications are developed; hard to define when an advantage is obtained.\\
\end{tabularx}
\caption{High level summary of current state-of-the-art for applications of quantum annealing using current hardware. \label{tab:applications}}
\end{table}

\subsection{Quantum simulation}\label{ssec:qsim}
Quantum simulation by using one quantum system to simulate another \citep[reviewed in][]{Georgescu2014RMP} is one of the most promising early uses of quantum computers where a quantum advantage is expected to be realised \citep{Daley2022NatPers}.  
A large class of quantum simulations consist of the preparation of a quantum state followed by analysis of the state using some variant of phase estimation \citep{Kitaev1996} to efficiently extract useful information from the quantum state.  Both preparation and phase estimation require the quantum state to be evolved using the relevant Hamiltonian.  In quantum simulators, the desired Hamiltonian evolution is implemented directly, as opposed to digitised.  Such quantum simulators are often referred to as ``special purpose'', because it is only necessary to implement one Hamiltonian, or class of Hamiltonians, that corresponds to the problem of interest.  Clearly, given a wide enough range of controls, quantum simulators could be used to simulate any reasonable physical quantum system \citep{Childs2010Ham}, and would be ``universal'' in at least a practical sense.  However, they are going to be most efficient for the Hamiltonians that best match their natural Hamiltonians.
As an example of directly using the quantum annealing hardware Hamiltonian, studies on D-Wave quantum annealers \citep{King2023SpinGlass,King2025beyondClassical} show how they can be used to explore the behaviour and phase diagram of large ($\sim 5000$) spin systems, sizes where classical simulation is difficult or unfeasible. These build on earlier spin-glass work \citep{Harris2018SpinGlass}. Additionally, the phase diagrams of a variety of other systems have been explored, for example, ice-like systems \citep{King2021SpinIce,Lopez-Bezanilla2023SpinIce}, one-dimensional spin-chains with coherent dynamics \citep{King2022Coherent}, spin-chain compound dynamics \citep{King2021SpinChainCompound}, fullerenes \citep{Lopezbezanilla2025Fullerines} and fractional magnetisation \citep{Kairys2020Fractional}. Non-equilibrium behaviours such as hysteresis \citep{Pelofske2025Hysteresis,Barrows2025Hystoresis} and tunneling range \citep{Chancellor2021SearchRange} have also been explored.
Ultracold atoms provide another route for analog simulation \citep{GrossQSimUltracold}, although such experiments are often configured for the simulations, rather than being an alternative use of a quantum annealer. These include studies of frustrated magnetism on triangular lattices \citep{Stuck2011Frustrated}, superfluidity effects \citep{Kulov2003Superfluid}, themal sampling \citep{Camino2025ColdAtom} and many-body localisation \citep{Choi2016ColdAtomMBL}.

\subsection{Ground state and sampling problems}

As well as directly simulating condensed matter systems, there are many problems in condensed matter physics that naturally map to optimisation or sampling problems on quantum annealing hardware, such as determining ground state configurations \citep{Camino2025Disordered,Gusev2023D-WaveCrystal}, sign problems in quantum Monte Carlo, quantum state preparation, out-of-equilibrium dynamics, and open quantum systems. Methods for quantum field theory simulation in a high energy physics context have been proposed \citep{Abel2021QFT} with potential for adaption to a condensed matter physics setting. Annealers have been shown to perform well at Monte-Carlo sampling as well \citep{Nelson2022Sampling,Sandt2023Sampling}, particularly in the low-temperature regime which is often difficult to access by traditional methods due to high rejection rates.

Quantum annealers are well suited to simulating condensed matter physics problems because they typically naturally share a lower, two- or three-dimensional connectivity.   This is in contrast to many classical optimisation problems where highly connected all-to-all graphs result from encoding the constraints.  A concrete example of the difference this makes was shown by \cite{Camino2025Disordered}, by comparing between a highly connected constraint-based method and a more natural chemical-potential-based method. 
The problem \citeauthor{Camino2025Disordered} solve is to determine the stable configurations of solid solutions (alloys) with small numbers of defects or vacancies in the lattices, at finite temperatures.
A key technique they develop is to not constrain the number of defects directly, but rather, to encode a chemical potential for the defects in the fields of the Ising model. This allowed much more efficient processing, as the graph of the problem given to the annealer is planar and therefore embeds efficiently on the hardware graph. Furthermore, the annealer is not used to find the single most optimal configuration, but instead used as a thermal sampler to find low-energy configurations, leveraging the thermal equilibrium behaviour within the quasistatic regime. 
The work by \citeauthor{Camino2025Disordered} is thus a good example of a hybrid quantum-classical \citep{Callison2022Hybrid} algorithm, using the annealer only for the sampling task it is good at, while the full workflow performs expensive classical DFT (density functional theory) calculations only for the low energy samples which the annealer was able to identify. 
They illustrate the effectiveness of the method by predicting bandgap bowing in $Al_{1-x}Ga_xN$, and bulk modulus variations in $Ta_{1-x}W_x$, with results in excellent agreement with experiments.  Current hardware is not yet able to address problems at the forefront of current research, such as high-entropy super alloys, but the potential for tackling useful problems in the near future is clear.

\section{Outlook}\label{sec:outlook}

There is still much to do to develop theory to fully understand the new features becoming available in quantum annealing hardware. The field has reached a crucial stage, with rapid quenches now available in flux qubit (D-Wave) devices and high coherence in neutral atom annealers, both of which should provide access to a highly coherent diabatic regime.
This is a step change from even the very recent past, when only quasistatic annealing was available. Intermediate scale, programmable, far from equilibrium quantum systems can now be used to test theory and explore new quantum many body behaviours. 
These new tools demand new physics to fully understand their behaviour, and hence how they can best be used for computational tasks.  Specifically, in the near term it would be useful to obtain better understanding of how system scaling to larger sizes affects the diabatic regime.

Questions still remain in the quasistatic regime, too, particularly in cases where a simple Kibble-Zurek picture fails, and in understanding how distributions containing traces of transverse fields can be used in practice \citep{chancellor2016DWaveKZ}.
Fully characterising the circumstances under which a quasistatic approximation is valid still needs to be done, and it would be useful to understand the limitations of annealers as thermal samplers. Another important open question is how the interference effects, which separate physical annealing from quantum Monte Carlo, manifest in real condensed matter systems, and under which circumstances these are important.  This will provide criteria for when quantum Monte Carlo can provide accurate results using classical computation, and when alternative methods are needed.

The success of quantum annealing has inspired a wide range of related algorithms, for example, the quantum approximate optimisation algorithm (QAOA) \citep{farhi2014QAOA} is directed inspired by quantum annealing. Instead of directly implementing the Hamiltonian, QAOA adapts it to a gate model setting with alternating applications of the driver and problem Hamiltonians, with a classical variational component to tune the alternating steps \citep{Callison2022Hybrid}. 
On the other end of the spectrum, quantum annealing has rekindled an interest in classical simulated annealing algorithms, and in specialised CMOS hardware for these algorithms, known as digital annealers \citep{Aramon2019DA}. 
Related approaches have also been tested in an optical setting, for example, combinatorial optimisation has become a target for variational boson sampling hardware \citep{bradler2021ORCA,goldsmith2024ORCA+TSP}, building on the seminal theory of \cite{Aaronson2011}. Other approaches directly use interference to encode optimisation in optical systems.  For example, coherent Ising machines based on nonlinear optical effects are able to perform amplification and encoding in the phase of coherent light \citep{Yamamoto2017CoherentIsing}. Entropy computing \citep{Nguyen2025entropycomputing} is a similar, but distinct, scheme.  Few of these have reached the stage of being available online for users to test their applications, but any hardware that can reduce the power consumption for equivalent computational outputs is of interest and relevance on the path towards net zero.

Despite the barriers to scaling for highly connected traditional optimisation problems, the (relatively) low dimensionality of condensed matter physics problems makes them well-suited to current hardware.  Annealers also seem to be better at approximate optimisation/sampling than exact optimisation, which again is easier to leverage for solving condensed matter physics problems.  Creative and encouraging examples have been appearing in the literature in the last few years, and there is much mutual benefit to be gained by closer collaboration between condensed matter physicists and quantum annealing researchers.

\ack{%
For the purpose of open access, the authors have applied a Creative Commons Attribution (CC BY) licence to any Author Accepted Manuscript version arising from this submission.}  

\funding{
VK and NC funded by UKRI EPSRC grants EP/T026715/2, EP/T001062/1, EP/W00772X/2, and EP/Y004566/1, EP/Z53318X/1, and the UKRI DRI and STFC funded CCP-QC Bridge Project -- extended case studies for neutral atom hardware. 
}

\addcontentsline{toc}{section}{References}

\end{document}